\documentstyle[espcrc2,fleqn,twoside,epsfig]{article}

\title{Superfluid Inhomogeneity and Microwave Absorption
in Model High-T$_c$ Superconductors}
\author{Sergey\ V.\ Barabash, David Stroud
\address{Department of Physics,
The Ohio State University, Columbus, Ohio 43210}
}

\begin{document}

\def\etal{ {\it et.\ al.}}
\def\eps{\epsilon}
\def\tnsr{\tensor}
\def\beq{\begin{equation}}
\def\eeq{\end{equation}}
\newcommand{\bscco}{{Bi$_2$Sr$_2$CaCu$_2$O$_{8+\delta}$} }

%\begin{frontmatter}

\begin{abstract}
\vspace{2pc}

We investigate the microwave absorption arising from inhomogeneity 
in the superfluid density of a model high-T$_c$ superconductor.  Such 
inhomogeneities may arise from a wide variety of
sources, including quenched random disorder and static charge 
density waves such as stripes.  We show that both mechanisms will 
inevitably produce additional absorption at finite frequencies.  
We present simple model calculations for this extra absorption, and 
discuss applications to other transport properties in high-T$_c$ 
materials.  Finally, we discuss the connection of these predictions 
to recent measurements by Corson\etal\cite{corson} of absorption 
by the high-temperature superconductor \bscco
in the THz frequency regime.

\vspace{5pc}
\end{abstract}
\maketitle

%\begin{keyword}
%\PACS
%\end{keyword}

%\end{frontmatter}

\section{Introduction}

The high-T$_c$ superconductor \linebreak 
\bscco  shows a remarkably strong
absorption in the microwave regime, even far below 
$T_c$\cite{corson,lee}.  In conventional,
low-T$_c$, s-wave superconductors, there is no such
background, because there is no absorption below 
the energy gap for pair excitations, $2\Delta$.  
But in high-T$_c$ materials, which are thought to have a 
$d_{x^2-y^2}$ order
parameter, the gap vanishes in certain nodal $k$ directions.  Hence,
gapless nodal quasiparticles can be excited, and hence
can absorb microwave radiation, at arbitrarily low
temperatures.  However, experiment suggests that this
absorption is much stronger than expected from the quasiparticles
alone\cite{corson2}.  

In this paper, we show that such extra absorption  
can be produced if the superfluid density $n({\bf x})$ within the 
$ab$ plane is a function of position ${\bf x}$.  
Such spatial variation can be produced, e. g., by quenched disorder,
or by charge density waves (e. g. stripes).  The extra absorption
is most likely in the microwave frequency range.  A similar
absorption was found by us in a previous paper\cite{barabash},
in which the inhomogeneity was described as static fluctuations
in the Josephson coupling between superconducting grains. 

\section{Microwave Absorption in an Inhomogeneous Two-Dimensional
Superfluid}

We consider a two-fluid model of a superconductor in
two dimensions, with local conductivity
\begin{equation}
\sigma({\bf x}, \omega) =  \sigma_{qp}(\omega) + 
\frac{iq^2n_s( \omega, {\bf x})}{m^*\omega},
\end{equation}
where $q = 2|e|$ and $m^*$ is twice the electron mass $m_e$.
$n_s({\bf x}, \omega)$ is assumed spatially varying, 
while $\sigma_{qp}$ is taken to be spatially uniform.  We wish to
calculate the complex effective conductivity $\sigma_e(\omega)$.  
$\sigma_{qp}$ might be a contribution from
the nodal quasiparticles, while $n_s$ represents the 
perfect-conductivity response of the superconductor.  
It can be spatially varying because of the very short
in-plane coherence length.

%CHANGE --I reworded the next sentence
We employ the 
Kramers-Kronig relations satisfied by
$\sigma_e(\omega)$, namely
$\sigma_{e1}(\omega) = \frac{2}{\pi}P\int_0^\infty
\frac{\omega^\prime \sigma_{e2}(\omega^\prime)}
{\omega^{\prime 2} - \omega^2}d\omega^\prime + \sigma_\infty$,
and
$\sigma_{e2} = \frac{q^2n_{s,e}}{m^*\omega} - \frac{2\omega}{\pi}
P\int_0^\infty\frac{\sigma_{e1}(\omega^\prime)-\sigma_\infty}
{\omega^{\prime 2} - \omega^2}d\omega^\prime$,
where P means ``principal part of,'' 
$\sigma_e = \sigma_{e1} + i\sigma_{e2}$, and 
$n_{s,e}$ is the effective superfluid density 
%CHANGE -- next sentence added
[the value of $\frac{q^2n_{s,e}}{m^*}$ can be found from the residue of
$\sigma_e(\omega)$ at $\omega=0$].
At very large frequencies $\omega$, the equation for $\sigma_{e2}$
becomes
$\sigma_{e2} \rightarrow \frac{q^2n_{s,e}}{m^*\omega} + 
\frac{2}{\pi\omega}\int_0^\infty
[\sigma_{e1}(\omega^\prime) - \sigma_\infty]d\omega^\prime$.

\subsection{Frequency-Independent $\sigma_{qp}$}

If $\sigma_{qp}(\omega)$ is real and  
frequency-independent, then  $\sigma_\infty = \sigma_{qp}$.   
Then at high frequencies, the {\em local} complex conductivity,
$\sigma_\infty + iq^2n_s({\bf x})/(m^*\omega)$, has only small 
spatial fluctuations. 
Then\cite{bergman} 
%CHANGE I changed \sim to \approx where appropriate throughout the text --SB
\begin{equation}
\sigma_e \approx \sigma_{av} -\frac{1}{2}\frac{\langle (\delta\sigma)^2\rangle}
{\sigma_{av}},
\label{sigma_e}
\end{equation}
$\delta\sigma(\omega, {\bf x}) \equiv \sigma(\omega, {\bf x}) -
\sigma_{av}(\omega)$, where
$\sigma_{av}(\omega) = \sigma_\infty + iq^2n_{s,av}/(m^*\omega)$,
$\delta\sigma(\omega, {\bf x}) = \sigma(\omega, {\bf x}) -
\sigma_{av}(\omega)$, and
$\langle...\rangle$ is a space average.   
Since only $n_s$, and not $\sigma_{qp}$, is fluctuating, 
this expression simplifies to 
%CHANGE numbered the formula and changed the next sentence
\beq
\sigma_e \approx \sigma_\infty + \frac{iq^2n_{s,av}}{m^*\omega} + 
\frac{1}{2}
%\left(
\frac{(q^2/m^*)^2\langle(\delta n_s)^2\rangle}
{\omega^2\sigma_\infty + iq^2\omega n_{s,av}/m^*}
%\right)
.
\label{eq:sigma_e_const}
\eeq
At large $\omega$, the imaginary part of $\sigma_e$
to leading order in $1/\omega$ is simply
$\sigma_{e2} \sim q^2n_{s,av}/(m^*\omega)$.
Equating this expression to the right-hand
side of the Kramers-Kronig expression for $\sigma_{e2}$,
we finally obtain
\begin{equation}
\int_0^\infty
\left[\sigma_{e1}(\omega^\prime)-\sigma_\infty\right]d\omega^\prime
= \frac{q^2\pi}{2m^*}\left(n_{s,av} - n_{s,e}\right)
\label{eq:sige1}
\end{equation}
If $n_s({\bf x})$ is spatially varying, the right-hand side
is always positive, whence 
{\em there will be an additional contribution to $\sigma_{e1}(\omega)$,
beyond $\sigma_{qp}$}.

%CHANGE
At small $\omega$, eq.\ (\ref{eq:sigma_e_const}) implies
$\sigma_{e2} \sim \frac{q^2n_{s,av}}{m^*\omega} + 
\frac{1}{2}\left(\frac{q^2/m^*\langle(\delta n_s)^2\rangle}
{\omega n_{s,av}}\right)$, and thus
$n_{s,av} - n_{s,e} \approx \frac{q^2n_{s,av}}{m^*}
\langle(\delta n_s)^2\rangle/(2n_{s,av}^2)$.
Then eq.\ (\ref{eq:sige1}) becomes
\begin{equation}
\int_0^\infty\left[\sigma_{e1}(\omega^\prime)-\sigma_\infty\right]
	d\omega^\prime
\approx \frac{q^2\pi n_{s,av}}{4m^*}\frac{(\delta n_s)^2}{n_{s,av}^2}.
\label{eq:sige1p}
\end{equation}
For fixed $\langle (\delta n_s)^2\rangle/n_{s,av}^2$,
this integral is proportional to $n_{s,av}$.
That is, the extra integrated fluctuation contribution 
to $\sigma_{e1}$, is proportional to the average 
superfluid density.  A similar result has been reported in 
%CHANGE added ref.
experiments\cite{corson,corson2}.

\subsection{Drude $\sigma_{qp}(\omega)$} 

For a Drude $\sigma_{qp}(\omega) = \sigma_0/(1-i\omega\tau)$,
we can carry out a similar analysis\cite{barabash1}.  
In the case of weak fluctuations in $n_s$, the
result is\cite{barabash1}
\begin{eqnarray}
\int_0^\infty &&\!\!\!\!\!\!\!\!\!\!\!\!
\left[\sigma_{e,1}(\omega^\prime)-\sigma_{qp,1}(\omega^\prime)
\right]d\omega^\prime \nonumber \\
&&  \!\!\!\!\!\!\!\!\!\!\! \approx 
\frac{\pi}{4}\left((q^2/m^*)^2\langle(\delta n_s)^2\rangle\right) \\
&& \!\!\times
\left(\frac{1}{q^2n_{s,av}/m^*}-\frac{1}{q^2n_{s,av}/m^*
+\sigma_0/\tau}\right). \nonumber
\end{eqnarray}

%CHANGE -typo
In the limit $\sigma_0/\tau \gg q^2n_{s,av}/m^*$, 
this expression reduces to the right-hand side of eq.\ (\ref{eq:sige1p}). 
Thus, in this regime, the extra spectral weight is indeed proportional to
the average superfluid density $n_{s,av}$.   

\subsection{Tensor $n_s$}

Next, we consider a {\em tensor} superfluid density, 
as would be expected in a superconducting layer containing 
quenched charge density waves such as a charge 
stripes\cite{kivelson}.  In this case, the ($2 \times 2$) 
superfluid density tensor should have the form
$n_s^{\alpha\beta}({\bf x}) = R^{-1}({\bf x})n_s^dR({\bf x})$,
where $n_s^d$ is a diagonal $2 \times 2$ matrix with 
diagonal components $n_{s,A}$, $n_{s,B}$, and $R({\bf x})$ 
is a position-dependent $2\times 2$ rotation matrix. describing
the relative orientation of the charge density wave or stripes.
If the stripes have either of two orientations
along the crystal axes of the layer, with equal probability,
then the {\em effective} superfluid density $n_{s,e}$ will
be a scalar.

The arguments of the previous subsections can readily be transferred to the
tensor case.  For a two-fluid model with
a frequency-independent quasiparticle conductivity which has
the same value $\sigma_{\infty}$ in both the $A$ and $B$ directions, 
the effective scalar conductivity $\sigma_e(\omega)$
again satisfies Kramers-Kronig relations, and
one again obtains the sum rule (\ref{eq:sige1}), where
$n_{s,av}$ is now the rotational average of a diagonal
element of $n_{s;\alpha\beta}$.  If the principal axes of
$n_{s;\alpha\beta}$ point with equal probability along the two symmetry
directions of the CuO$_2$ plane, as mentioned above, 
$n_{s,av} = (n_{s,A} + n_{s,B})/2$.  Likewise, if the principal axes 
were to point in any direction in the
plane with equal probability (a circumstance which seems unlikely for a
stripe phase), then it can be shown\cite{kazaryan} that
once again $n_{s,av} = (n_{s,A} + n_{s,B})/2$.

If the quasiparticle conductivity is frequency-dependent, then 
the analogous scalar results
of the previous section continue to hold.  For example, if 
$\sigma_{qp,A}(\omega)=\sigma_{qp,B}(\omega)$, then the extra 
spectral weight
due to the superfluid inhomogeneity is again given by 
eq.\ (5) in the weak-inhomogeneity regime.

In the case of the stripe geometry, where the 
principal axes of the conductivity tensor take either of
two perpendicular orientations with equal probability, 
$n_{s,e}$ is given by the duality result\cite{bergman,kivelson}
\begin{equation}
n_{s,e} = \sqrt{n_{s,A}n_{s,B}}.
\end{equation}
This form allows the extra spectral weight to be evaluated 
straightforwardly, given $\sigma_{qp}(\omega)$,
without making the small fluctuation approximation.

\section{Numerical Example}

\begin{figure*}[hbt]
\epsfig{file=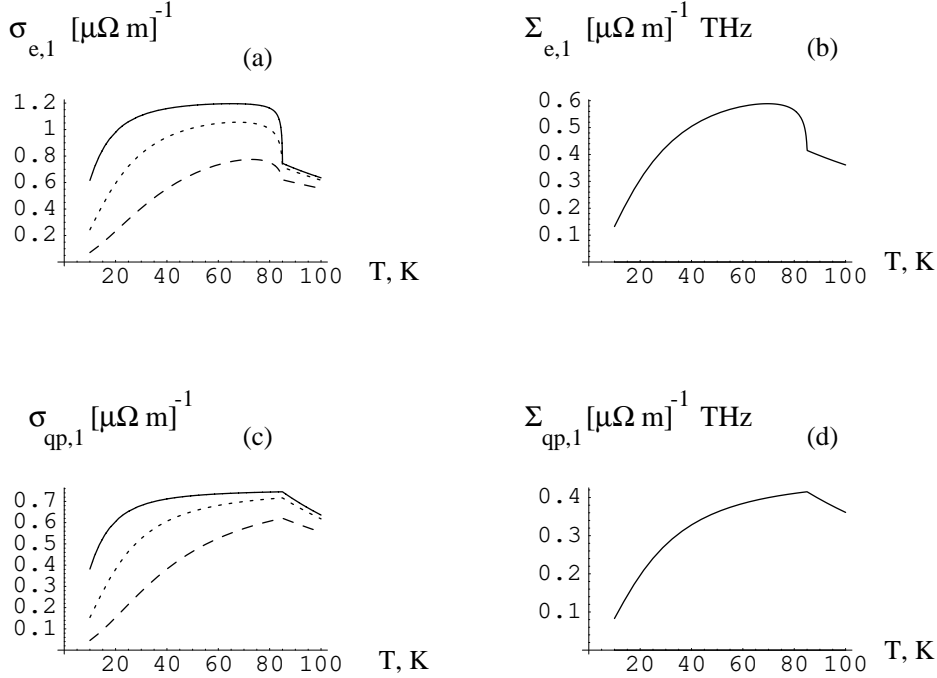, width=5in}
\caption{
(a) $\sigma_{e,1}(\omega, T)$, 
for $\omega/(2\pi) = 0.2$ THz (solid line), $0.4$ THz (dotted line), 
and $0.8$ THz (dashed line), 
for the model inhomogeneous superconductor described in the text.  
Also plotted are 
(b) $\Sigma_{e,1}\equiv
\int_{\omega_{min}}^{\omega_{max}}\sigma_{e,1}(\omega)d\omega$,
(c) $\sigma_{qp,1}(\omega,T)$,
and (d) $\Sigma_{qp,1}\equiv
\int_{\omega_{min}}^{\omega_{max}}\sigma_{qp,1}(\omega)d\omega$, 
where $\omega_{min}/(2\pi) = 0.2$Thz and 
$\omega_{max}/(2\pi) = 0.8$THz.
}
\end{figure*}

As a simple example, we consider a superconducting layer 
in which the conductivity has one of two possible values,
$\sigma_A(\omega)$ or $\sigma_B(\omega)$ with equal probability,
and we assume
$\sigma_{A,B}(\omega) =  \sigma_{qp}(\omega) + \frac{q^2n_{s;A,B}}
{m^*\omega}$, 
where we take $n_{s,A} > n_{s,B}$.  
This model would be suitable either for a layer with static
scalar disorder, or for a model of stripe domains, as discussed
above.  We also write
$\sigma_{qp} = \sigma_0/(1-i\omega\tau_{qp})$, with
$\sigma_0 = n_{qp}q^2\tau_{qp}/m^*$.
We also assume $n_{qp} = \alpha T$ for $T<T_c$ or
$n_{qp} = \alpha T_c$ for $T>T_c$,
$\tau_{qp}^{-1} = \beta T$, $n_{s,A} = \gamma n_{s,0}$,
$n_{s,B} = \gamma^{-1} n_{s,0}$, where 
$n_{s,0}(T) = n_{s,0}(0)\sqrt{1-2\alpha T/n_{s,0}(0)}$, and
%CHANGE
$\gamma$ is a parameter describing the superfluid inhomogeneity
(this form ensures that 
$n_{s,av}\equiv\sqrt{n_{s,A}n_{s,B}}=n_{s,0}$).  
Our choice of temperature dependence for $n_{s,0}(T)$ ensures that $n_{s,0}$
(i) decreases linearly with increasing temperature $T$ at small $T$,
as observed experimentally\cite{linear}; and (ii) vanishes at
a critical temperature $T_c$ as $\sqrt{T_c - T}$.
The ingredients of this model are very similar to those 
of Ref.\ \cite{corson2}, and have a straightforward interpretation.
First, $n_{s,qp} $ should be 
proportional to $T$ in a gapless $d$-wave superconductor\cite{gap},
and hence $n_{s,e}$ should be depleted by the same amount
and fall off linearly in $T$.  We also assume
that $n_{s,A}$ and $n_{s,B}$ individually are linear in $T$.
The form $1/\tau_{qp} = \beta T$, 
where $\beta$ is another constant\cite{corson2},
has been observed for nodal quasiparticles in the superconducting
states\cite{qp}.

We compute $\sigma_e$ using Bruggeman effective-medium approximation 
(EMA)\cite{bergman,bruggeman}, which gives
$(\sigma_A-\sigma_e)/(\sigma_A+\sigma_e)
+ (\sigma_B-\sigma_e)/(\sigma_B+\sigma_e) = 0$.
The solution to this equation is simply
$\sigma_e = \sqrt{\sigma_A\sigma_B}$.
Fig.\ 1 shows the resulting $\sigma_{e,1}(\omega, T)$ for several
frequencies ranging from 0.2 to 0.8 THz, the range measured in Ref.\
\cite{corson2}.  The parameters $T_c$, $\sigma_0$, 
$n_{s,0}(0)$, $\alpha$ and $\beta$ were taken from 
Ref.\cite{corson2}, and we assumed $\gamma=3$.
Also shown are 
$\int_{\omega_{min}}^{\omega_{max}}\sigma_{e,1}(\omega,T)d\omega$ 
for $\omega_{min}/(2\pi) = 0.2$THz and $\omega_{max}/(2\pi) = 
0.8$THz.  Finally, we plot
$\sigma_{qp,1}(\omega, T)$ for these frequencies,
as well as
$\int_{\omega_{min}}^{\omega_{max}}\sigma_{qp,1}d\omega$.
%CHANGE  I assume we do not need $n_{s,e}(T)$ -- what do you think?
%and $n_{s,e}(T)= (m^*\omega/q^2)\sigma_{e,2}(\omega)$.
Clearly, $\sigma_{e,1}(\omega)$ is considerably increased 
beyond the quasiparticle contribution, because of 
spatial fluctuations in the superfluid density.  
%However, the
%integrated fluctuation conductivity
%$\int_{\omega_{min}}^{\omega_{max}}$
%Re$[\sigma_{e,1}(\omega,T)-\sigma_{qp}(\omega, T)d\omega$ is not
%proportional to the superfluid density $n_s(T)$, in contrast to experiment.

\section{Discussion}

We have shown that a superconducting layer with an 
inhomogeneous superfluid density will have an extra absorption 
not present in a homogeneous superconductor.
The frequency integral of $\sigma_{e,1}(\omega)$
associated with this absorption is 
proportional to the superfluid density, 
in general agreement with experiments\cite{corson,corson2}.  
We have also shown that this inhomogeneity, and hence the
extra absorption seen in experiments, can arise from
a stripe domain structure, as well as from random but isotropic
disorder.  Thus, we speculate that no such extra absorption should
be observed in a high-T$_c$ superconductor unless one of these
two types of inhomogeneities are present (beyond that expected
purely from nodal quasiparticle absorption).

\section{Acknowledgments.} This work has been supported by NSF Grant
DMR01-04987, and by the U.-S./Israel Binational Science Foundation.

\end{document}